\documentclass[twocolumn]{aastex631}

\usepackage{physics}

\begin{document}

\title{Mass Ejection Driven by Sudden Energy Deposition in Stellar Envelopes}

\author[0000-0002-0088-2563]{Nicholas J. Corso}
\affiliation{Department of Astronomy, Center for Astrophysics and Planetary Science, Cornell University, Ithaca, NY 14853, USA}

\author[0000-0002-1934-6250]{Dong Lai}
\affiliation{Department of Astronomy, Center for Astrophysics and Planetary Science, Cornell University, Ithaca, NY 14853, USA}
\affiliation{Tsung-Dao Lee Institute, Shanghai Jiao-Tong University, Shanghai,  China}

\begin{abstract}

A number of stellar astrophysical phenomena, such as tidal novae and planetary engulfment, involve sudden injection of sub-binding energy in a thin layer within the star, leading to mass ejection of the stellar envelope. We use a 1D hydrodynamical model to survey the stellar response and mass loss for various amounts ($E_{\mathrm{dep}}$) and locations of the energy deposition. We find that the total mass ejection has a nontrivial dependence on $E_{\mathrm{dep}}$ due to the varying strengths of mass ejection events, which are associated with density/pressure waves breaking out from the stellar surface. The rapid occurrence of multiple breakouts may present a unique observational signature for sudden envelope heating events in stars.

\end{abstract}

\section{Introduction} \label{sec:intro}

Mass ejection from a star resulting from sudden energy infusion is a common occurrence during stellar evolution. A well-known example is the thermal instability associated with shell Helium burning (``He-shell flashes'') in AGB stars, leading to thermal relaxation oscillations of the stellar envelope and the associated mass loss \citep{Schwarzschild&Harm65}. Another example is classical novae, luminous eruptions associated with white dwarfs accreting matter from nondegenerate binary companions \citep{Gallagher&Starrfield78, Chomiuk+21}. Under appropriate conditions (e.g. a certain range of accretion rates), the accreted layer undergoes unstable (runaway) nuclear burning upon reaching a critical mass, and the resulting energy release causes the accreted envelope to expand, ultimately leading to its ejection.

Enhanced mass loss due to energy deposition in stellar envelopes has also been discussed in the context of massive stars prior to supernova explosion.  Such elevated mass loss may help explain the luminosity evolution and variability of Type IIn SNe as a result of the interaction between SN ejecta and circumstellar material. \cite{Quataert&Shiode12} (also \citealt{Shiode&Quataert14}) suggest that vigorous core convection in pre-SN stars excite internal gravity waves that leak out as acoustic waves in the stellar envelope, and the dissipation of the waves then leads to mass ejection of the outer layer and pre-SN outbursts.  A number of follow-up works have examined various aspects of this wave-heating scenario \citep{Fuller17, Fuller&Ro18, Leung+21a, Leung+21b, Wu&Fuller21}. Energy deposition by a companion during common envelope evolution may be an alternative mechanism of driving the required mass loss \citep{Chevalier12}. The response of the stellar envelope to a steady energy injection has been studied by \cite{Quataert+16}, \cite{Kuriyama&Shigeyama20}, and \cite{Ko+22}.

Sudden energy release in stellar envelopes can also occur in other contexts. One example is ``tidal novae'' in coalescing binary white dwarfs prior to merger \citep{Fuller&Lai12}. In these systems, the time-varying tidal force from the companion excites internal gravity waves, which then propagate outward and dissipate in the outer envelope of the white dwarf (e.g. \citealt{Su+20}). The heating rate increases rapidly as the binary orbit decays, and at some point the heating becomes significant enough to ignite hydrogen burning, creating a thermonuclear runaway that releases even more energy. Another very different example involves an evolved star engulfing an orbiting giant planet or brown dwarf \citep{OConnor+23}. As the planet spirals inward due to gas drag, it plows through the stellar envelope and releases energy. As the inspiral speeds up, the largest amount of energy release occurs somewhat suddenly, just before the planet ``dissolves'' (e.g., by the tidal force from the stellar core).

\begin{figure*}[ht]
\begin{center}
\includegraphics[width=\textwidth]{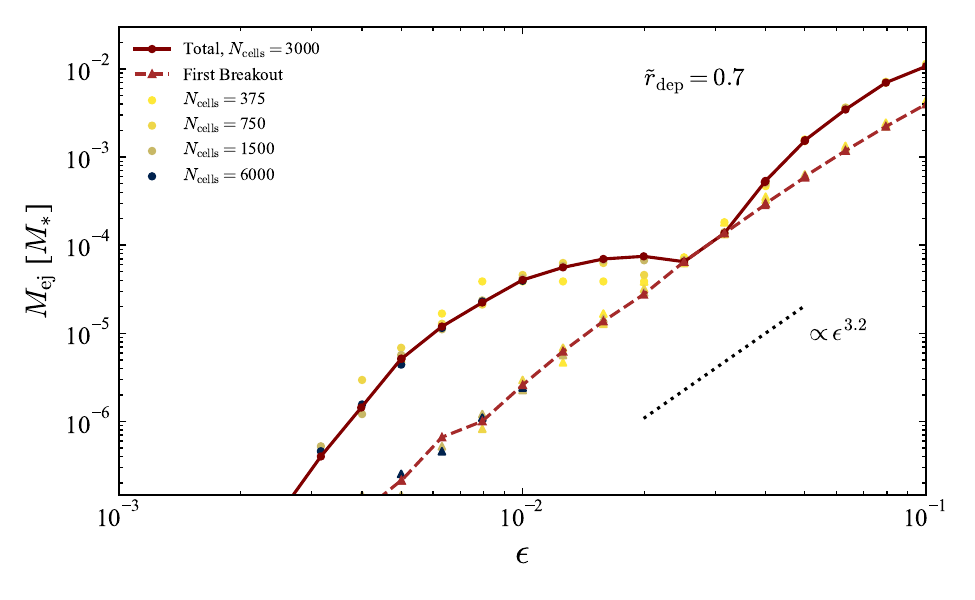}
\caption{Ejected mass as a function of the dimensionless energy deposition, $\epsilon=E_\mathrm{dep} R_\ast / (GM_\ast^2)$ for the deposition radius $\tilde{r}_\mathrm{dep}=0.7$. The solid line with circular points represents the total ejected mass, while the dashed line with triangular points depicts only the mass ejected from the first breakout (measured at $t=2.5t_\mathrm{dyn}$). The separation between the two lines indicates the presence of multiple mass breakout events. Our fiducial resolution has $N_\mathrm{cells}=3000$ (maroon). The dots of different color depicts results at different resolutions. Also included is a dotted line of spectral slope $3.2$, representing the scaling of the mass loss in the first breakout with $\epsilon$. \label{massej}}
\end{center}
\end{figure*}

All of the above examples share the common feature in that a certain amount of energy is released quickly for a short period of time in a narrow region of the stellar envelope. The amount of energy release and the location of heat deposition depend on the specifics of the problem. Some previous works have explored the response of the star (particularly mass loss) as a function of the amount and radial location of energy deposition, including \cite{Dessart+10} and \cite{Quataert+16}. However, the former generally concentrates on the energy deposition close to the core with an amount comparable to the binding energy of the star. The latter, meanwhile, focuses on very shallow energy depositions occurring over a relatively long timescale, such that a quasi-steady state is achieved, leading to a mostly hydrostatic stellar expansion with a strong wind.

In this study, we study the idealized problem where we assume that a sub-binding energy deposition occurs instantaneously in a narrow region of the star. This generally requires that the energy injection timescale be shorter than the local dynamical time. We determine the total mass loss as a function of the amount and radial location of the energy deposition. Using 1D hydrodynamics while ignoring the effects of radiation transport, we hope to obtain the ``base result'' on how mass loss scales with the energy deposition, and lay the groundwork for future explorations where additional physical effects can be added.

\newpage

\section{Problem Setup and Method} \label{sec:method}

Our initial star is a polytrope (satisfying the equation of state $P\propto\rho^\gamma$) with mass $M_\ast$ and radius $R_\ast$. We discretize its radial profile using cells of equal steps in radius. These correspond to non-uniform Lagrangian mass cells that are used in the ensuing hydrodynamical simulations. Unless stated otherwise, we use 3000 cells. We choose $\gamma=5/3$ and express mass and radius in units of $M_\ast$ and $R_\ast$.

\begin{figure*}[ht]
\begin{center}
\includegraphics[width=0.95\textwidth]{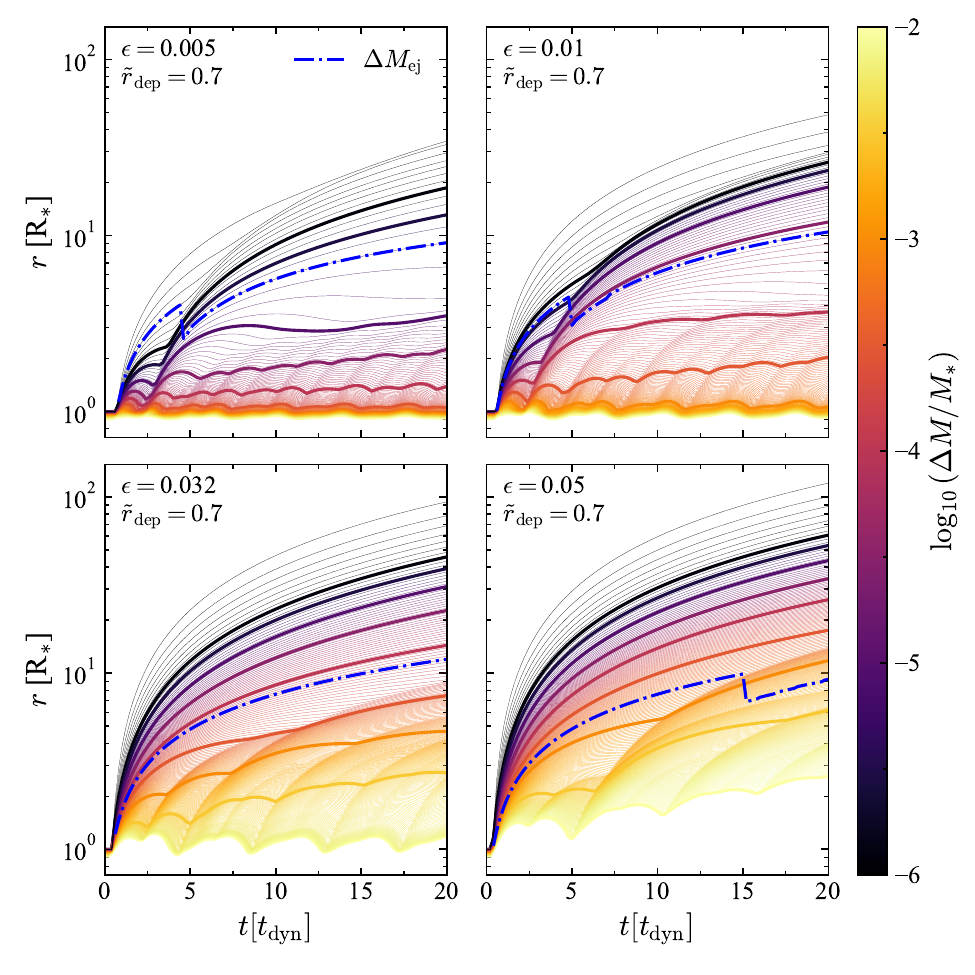}
\caption{Evolution of the radii of the outermost mass shells as a function of $t$ for energy injection $\epsilon=0.005, 0.01, 0.032, 0.05$ (left to right, top to bottom), all with $\tilde{r}_\mathrm{dep}=0.7$. The evolution is shown up to $t=20 t_\mathrm{dyn}$ to concentrate on the behavior of the second breakout, even though all simulations were run until at least $t=45 t_\mathrm{dyn}$. All cells with external mass $\Delta M \leq 10^{-2} M_\ast$ are shown, but the cells at half logarithmic steps in $\Delta M$ are represented with thicker lines for ease of reference. In each plot, the thick blue dash-dotted line represents the transition from bound to unbound material, such that all mass external to the cell is considered ejected. \label{radiusev}}
\end{center}
\end{figure*}

We simulate the ``instantaneous'' heat deposition in the star by increasing the specific internal energy of the gas in a thin layer. We specify this deposition using two dimensionless parameters. Namely, we deposit a total energy of $E_\mathrm{dep}=\epsilon \left(GM_\ast^2/R_\ast\right)$ in a layer with the radius range $r \in [\tilde{r}_\mathrm{dep}-\Delta\tilde{r}_\mathrm{dep}/2, \tilde{r}_\mathrm{dep}+\Delta\tilde{r}_\mathrm{dep}/2]R_\ast$. Throughout this study, we choose the width of the deposition layer to be $\Delta\tilde{r}_\mathrm{dep}=0.05$. We consider the dimensionless energy in the range $\epsilon\in[5\times10^{-4},10^{-1}]$ and the deposition radius in the range $\tilde{r}_\mathrm{dep}\in[0.6,1)$. Note that by raising the energy in the deposition layer instantaneously, we require that the energy is injected on a timescale shorter than the dynamical time, $t_\mathrm{dyn}=\left(R_\ast^3 / GM_\ast\right)^{1/2}$.

We run all simulations using a 1D first-order Lagrangian hydrodynamics code constructed using the scheme described by \cite{Bodenheimer+06} with the inclusion of self-gravity and artificial viscosity.
All simulations last for at least $45t_\mathrm{dyn}$. We choose this duration to provide sufficient time for the density/pressure waves generated by the initial energy injection and reflected by the stellar core and surface to affect the energetics of the stellar envelope. In some cases, it is necessary to run for a longer duration (up to $\sim150t_\mathrm{dyn}$) in order for the flow to reach a quasi-steady state and the mass ejection to be reliably determined.

To determine the mass ejection following the energy injection, we consider the Bernoulli parameter for each mass shell,
\begin{equation}
B = \frac{1}{2}v^2 + \Phi_\mathrm{g} + \left(\frac{\gamma}{\gamma-1}\right) \frac{P}{\rho},
\end{equation}
where $\Phi_\mathrm{g}$ is the gravitational potential and $v$ is the radial velocity. Note that $B$ is constant for barotropic flows in steady state; this is generically not satisfied for our system after the energy deposition. However, if the simulation is run for a sufficiently long time and the outflowing mass has reached a steady state, the ejected material approximately settles into a barotropic structure with no shock dissipation. As such, we measure $B$ across the profile for every simulation timestep, and after the system has reached its quasi-steady state, we consider a surface layer with $B>0$ to be unbound and therefore ejected.

\begin{figure}[ht]
\begin{center}
\includegraphics[width=0.5\textwidth]{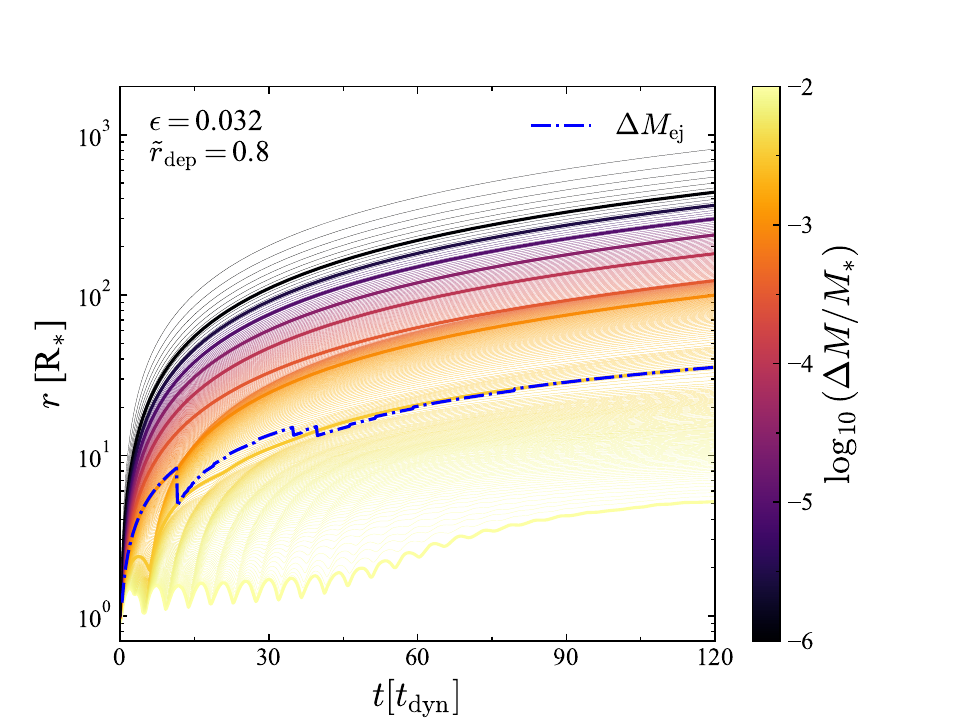}
\caption{Same as Figure \ref{radiusev} for $\epsilon=3.2\times10^{-2}$ and $\tilde{r}_\mathrm{dep}=0.8$. In this case, waves close to the unbound layer prevent a quasi-steady state from forming at $t \lesssim 45 t_\mathrm{dyn}$, and an accurate determination of the ejected mass requires long-term integration. \label{radiusev_long}}
\end{center}
\end{figure}

\section{Results} \label{sec:results}

\subsection{Fiducial Energy Deposition Radius}

Our fiducial runs assume the energy deposition radius $\Tilde{r}_\mathrm{dep}=0.7$. Figure \ref{massej} shows the mass ejection, $M_\mathrm{ej}$, as a function of $\epsilon=E_\mathrm{dep} / \left(GM_\ast^2/R_\ast\right)$ for our fiducial runs (solid line). While $M_\mathrm{ej}$ generally increases with $\epsilon$, a nontrivial ``wavy'' behavior emerges in the considered domain, revealing critical $\epsilon$ regions where the trend falls off, only to quickly revive.
Included in this figure is another curve labeled ``First Breakout,'' which shows $M_\mathrm{ej}$ measured at $t=2.3t_\mathrm{dyn}$. In our simulations, distinct mass ejection events (hereby termed ``breakouts'') occur, and $t=2.3t_\mathrm{dyn}$ generally divides the first and second breakouts, allowing us to observe the former separately. Despite the irregularity in the total $M_\mathrm{ej}$ vs. $\epsilon$ relation, the mass loss from the first breakout, $M_\mathrm{ej}^\mathrm{(first)}$, smoothly follows the power law, $M_\mathrm{ej}^\mathrm{(first)}\propto \epsilon^{3.2}$.
Figure \ref{massej} also shows the results from the simulations with different resolutions (characterized by the number of radial cells, $N_\mathrm{cells}$), indicating the level of convergence of our results.

We contextualize the nontrivial trend of the $M_\mathrm{ej}$ vs. $\epsilon$ relation using Figure \ref{radiusev}, which depicts the time evolution of the radii of the outermost mass shells until $t=20t_\mathrm{dyn}$ for a variety of test values of $\epsilon$. By tracing the cells of unbound material, we see that in each case the initial wave breakout ejects some mass. In most cases, a second wave breakout ejects more mass. However, in the other, often more energetic, ones, waves continually disturb the region surrounding the transition from bound to unbound material, preventing it from settling into a quasi-steady state. We ran these simulations, such as the one depicted in Figure \ref{radiusev_long}, for much longer durations, upwards of $t=120 t_\mathrm{dyn}$, until we achieved a quasi-steady state. In this example, we identify more than two breakout events. We observe this behavior for $\epsilon$ close to the critical regions, where the energy deposition is distributed more evenly among the rebounded material.

\begin{figure*}[ht]
\begin{center}
\includegraphics[width=0.95\textwidth]{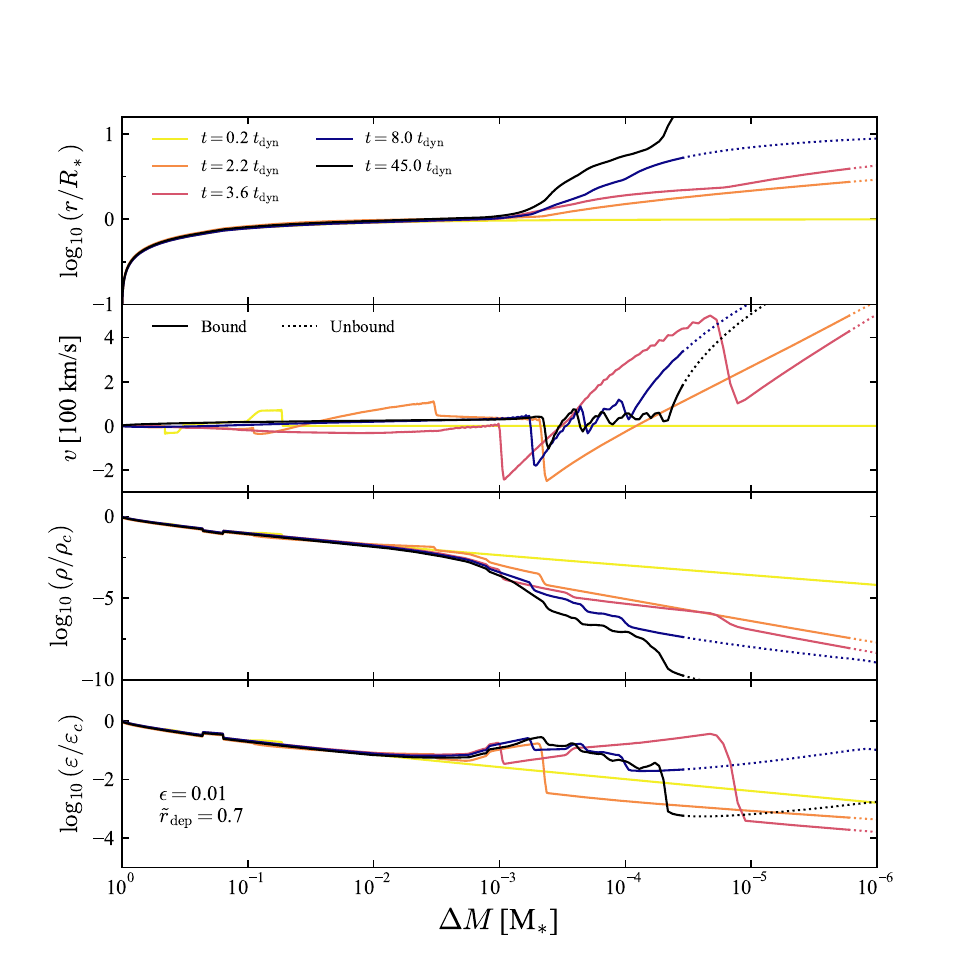}
\caption{Profiles of radius, radial velocity, density, and specific internal energy (top to bottom) as a function of the external mass at time steps $t=0.2, 2.2, 3.6, 8.0, 45.0 t_\mathrm{dyn}$ (yellow, orange, red, purple, black) for our fiducial run with $\epsilon=10^{-2}$ and $\tilde{r}_\mathrm{dep}=0.7$. The specific time steps are chosen to highlight the key stages in the time evolution of the system. In order, they are: (i) the initial generation of the ingoing and outgoing density waves; (ii) the rebound of the initially ingoing wave after the initially outgoing wave has ejected material; (iii) the generation of the shock of the second breakout as material that failed to eject during the first breakout crashes back down onto the stellar surface; (iv) the escape of the second breakout and beginning of pulsations in the remnant star; (v) the weakening of remnant oscillations in the underdense region. Bound material is represented by solid lines, while unbound material is represented by dotted lines. \label{profiles}}
\end{center}
\end{figure*}

\begin{figure*}[ht]
\begin{center}
\includegraphics[width=0.95\textwidth]{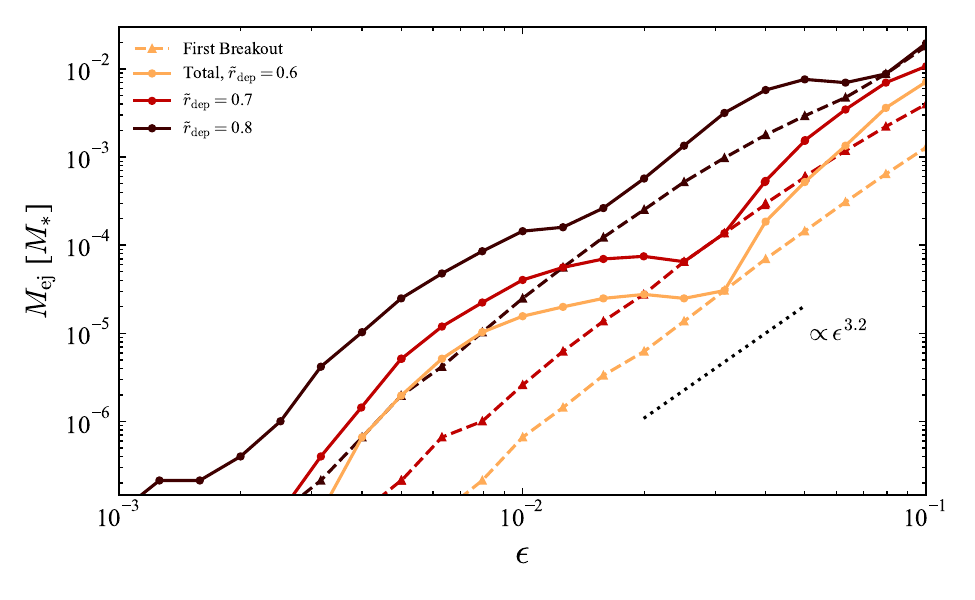}
\caption{Ejected mass as a function of $\epsilon$ for $\tilde{r}_\mathrm{dep}=0.6, 0.7, 0.8$ (bright to dark in increasing $\tilde{r}_\mathrm{dep}$). As in Figure \ref{massej}, the total mass ejected (solid line, circular points) is distinguished from the mass ejected by the first breakout alone (dashed line, triangular points), and the first breakout is described by a power law of spectral slope 3.2. \label{massej_r678}}
\end{center}
\end{figure*}

By comparing different values of $\epsilon$, the trend introduced in Figure \ref{massej} comes into focus. Most importantly, we see that at the critical values of $\epsilon$ with less ejected material, rebounding waves within the star do not contain enough energy to push through the unbound material and create a second breakout. As $\epsilon$ approaches these critical regions, the delay between the first and second breakouts increases. By contrast, at the $\epsilon$ values where the ejected mass locally peaks, the delay between the first and second breakouts is minimized. In other words, the timing of this delay appears to be tightly related to how much mass is ejected in the second breakout. We also note that above the critical region located at $\epsilon\simeq0.03$, more than one wave rebound is required to produce the second breakout.

Figure \ref{profiles} depicts the radius, radial velocity, density, and specific internal energy profiles as a function of $\Delta M$ (the amount of mass external to a shell) for the simulation initiated using our fiducial energy and deposition radius. We choose five key moments in the simulation to paint a more detailed picture of the stellar response to the heat deposition. We see that the energy deposition produces ingoing and outgoing density/pressure waves. The outgoing wave directly ejects the outer layers, causing the first breakout, while the ingoing wave reflects off the stellar core before proceeding to the surface to eject more material, causing the second breakout. Although much of the energy is carried off by unbound material from these two breakouts, matter that fails to unbind falls back to produce more ingoing waves, and these waves will approach the surface again after reflecting off the core. However, by this point, they have dissipated enough energy that they are no longer able to unbind more material from the star. Instead, they continuously reflect, causing the star to pulsate on a dynamical timescale. Remaining at the site of the original heat deposition is a remnant hot bubble, which had rapidly expanded at the start of the simulation. Additionally, a hot extended atmosphere lingers, composed of material just beneath the ejected layer.

\begin{figure*}[ht]
\begin{center}
\includegraphics[width=0.95\textwidth]{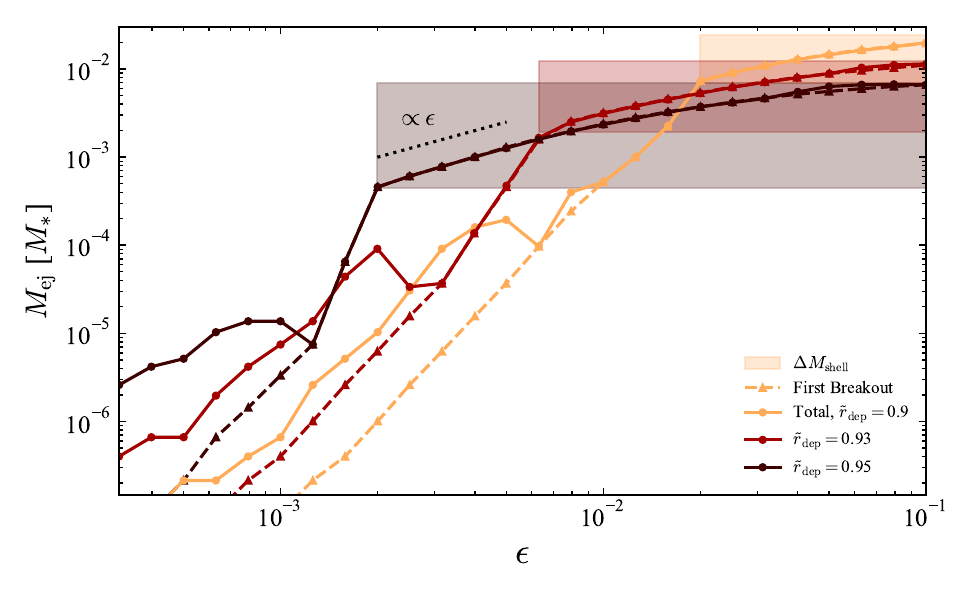}
\caption{Same as Figure \ref{massej_r678}, but for $\tilde{r}_\mathrm{dep}=0.9, 0.93, 0.95$. The shaded boxes correspond to the region of $\Delta M$ at the initial heat deposition, with the $\epsilon$ boundaries constrained to highlight its correspondence with the `high-energy' regime. Upon reaching this $\Delta M$ region, the slope of the curve flattens, with the black dotted line of spectral slope $1$ representing the new scaling.\label{massej_r995}}
\end{center}
\end{figure*}

\subsection{$\tilde{r}_\mathrm{dep}$-Dependence}

In Figure \ref{massej_r678} we repeat the analysis performed in Figure \ref{massej}, now for the energy deposition radius $\tilde{r}_\mathrm{dep}=0.6,0.7,0.8$. The three curves follow a similar trend, exhibiting the nontrivial ``wavy'' behavior shown in Figure \ref{massej}. However, for $\tilde{r}_\mathrm{dep}=0.8$, we note that the wavy behavior is weaker compared to the other cases, and the second breakout always occurs.
It should also be noted that in some cases, often with higher $\epsilon$ values, more than two breakouts may occur. As in Figure \ref{massej}, the mass ejected from the first breakout exhibits a rather consistent power law dependence on $\epsilon$, i.e. $M_\mathrm{ej}\propto\epsilon^{3.2}$. The steepness of this slope arises both from the direct energy increase as well as the increasing capability of the injected energy to reach directly to the surface to unbind material rather than distribute to the rest of the star.

Figure \ref{massej_r995} again repeats this analysis, now for larger $\Tilde{r}_\mathrm{dep}$ (energy deposition closer to the stellar surface). In doing so, we reveal a new, `high-energy' regime. Specifically, if the energy deposited is sufficiently large, the first breakout may eject most of the mass that had contained the original heat deposition layer. As a result, the remnant bubble at this layer may expand supersonically, forming a shock. If the energy injection is larger still, this shock, or even part of the heated remnant itself, may efficiently escape, causing the amount of ejected mass to noticeably increase. Afterward, the internal waves struggle to eject more material since the envelope has expanded so much, suppressing the second breakout almost entirely. The strong power law dependence of the first breakout mass ejection on $\epsilon$ transforms to a modest $M_\mathrm{ej}^\mathrm{(first)}\propto\epsilon$, and this slope decreases still at even higher values of $\epsilon$. Thus, in this `high-energy' regime, the first breakout `saturates,' such that increasing the deposited energy mostly energizes material that is already able to unbind at lower energies. Figure \ref{massej_r995} highlights this by showing the mass external to the inner radius of the heat deposition layer. In the `high-energy' regime, the curves asymptote to this mass, indicating that the shell fully escapes without any interior material.

\section{Discussion and Conclusions} \label{sec:discuss}

We have used 1D hydrodynamical simulations to model the evolution of a star following a sudden energy deposition in the stellar interior, to determine the stellar mass loss as a function of the amount ($E_{\mathrm{dep}}$) and location ($\tilde{r}_{\mathrm{dep}}$) of the energy deposition. We find that the ejected mass generally has a nontrivial dependence on $E_{\mathrm{dep}}$ (see Figs.~5 and 6). The mass ejection is associated with the breakouts of density/pressure waves that are generated by the energy deposition and the ensuing wave reflections from the stellar core and surface. While the mass ejection from the first wave breakout has a simple power-law dependence on the energy deposition ($M_{\mathrm{ej}}^{(\mathrm{first})} \propto E_{\mathrm{dep}}^{3.2}$), the total mass loss generally exhibits a wavy behavior due to the breakout of a second wave reflected off the stellar core. The strength of the second breakout depends on the time delay between the two breakouts, and in some cases there may be no second breakout.  For sufficiently large energy deposition close to the stellar surface, the envelope exterior to the deposition radius is efficiently ejected, and the total mass loss has a much weaker dependence on $E_{\mathrm{dep}}$.

The nontrivial dependence of $M_{\mathrm{ej}}$ on $E_{\mathrm{dep}}$ implies that in astrophysical situations involving sudden energy deposition (such as tidal nova and planet engulfment), one should exercise caution when trying to infer the total energetics of the event from observations. The multiple mass breakouts may be a unique feature of the event. Of course, since our study does not include any treatment of radiation, further works are needed to clarify the observational signatures of such energy deposition events.

Beyond the inclusion of radiation physics, there are a number of extensions of our work. First, we have adopted a simplified stellar model (i.e. polytrope) in this paper, and it will be useful to consider more realistic stellar models in future works. Second, we have assumed sudden energy injection in this paper. It will be important to consider spreading the energy injection over a finite duration of time, which may result in a more gradual mass ejection (see Quataert et al.~2016). Extending the duration may blur the distinction between the multiple mass breakout events.  Finally, we have focused on 1D flows in this paper. Expanding the simulations to 2D or 3D would help to address anisotropies that may emerge in the system, particularly in the situations where the energy deposition is not spherically symmetric.

\begin{acknowledgments}
This research was partially supported by the NSF grants AST-2107796 and DGE-2139899.
\end{acknowledgments}

\bibliography{Bibliography}{}
\bibliographystyle{aasjournal}

\end{document}